\documentclass[aps,prl,reprint,superscriptaddress,floatfix]{revtex4-2}
\usepackage{graphicx}
\usepackage{dcolumn}
\usepackage{bm}
\usepackage{amsmath}
\usepackage{overpic}
\usepackage{subfigure}
\usepackage{ragged2e}
\usepackage{amssymb}

\usepackage[colorlinks=true,citecolor=blue,urlcolor=blue]{hyperref}

\DeclareMathOperator{\sech}{sech}
\usepackage{physics}
\usepackage{tikz}
\usepackage{verbatim}

\newcommand{\subfigref}[2]{%
  Fig.~\textcolor{blue}{\ref{#1}(#2)}%
}

\begin{document}

\title{Broadband Population Transfer Based on Suture Adiabatic Pulses}
%\title{High fidelity broadband population transfer based on parallel chirped pulses method}% Force line breaks with \\

\author{Jiaming Li}
%\affiliation{Laboratory of Quantum Information, University of Science and Technology of China, Hefei 230026, China}
\affiliation{Anhui Province Key Laboratory of Quantum Network, University of Science and Technology of China, Hefei 230026, China}
\affiliation{CAS Center For Excellence in Quantum Information and Quantum Physics, University of Science and Technology of China, Hefei 230026, China}

\author{Xi-Wang Luo}\email{luoxw@ustc.edu.cn}
%\affiliation{Laboratory of Quantum Information, University of Science and Technology of China, Hefei 230026, China}
\affiliation{Anhui Province Key Laboratory of Quantum Network, University of Science and Technology of China, Hefei 230026, China}
\affiliation{CAS Center For Excellence in Quantum Information and Quantum Physics, University of Science and Technology of China, Hefei 230026, China}
\affiliation{Hefei National Laboratory, University of Science and Technology of China, Hefei 230088, China}
\affiliation{Anhui Center for Fundamental Sciences in Theoretical Physics,
University of Science and Technology of China, Hefei 230026, China}

\author{Guang-Can Guo}
%\affiliation{Laboratory of Quantum Information, University of Science and Technology of China, Hefei 230026, China}
\affiliation{Anhui Province Key Laboratory of Quantum Network, University of Science and Technology of China, Hefei 230026, China}
\affiliation{CAS Center For Excellence in Quantum Information and Quantum Physics, University of Science and Technology of China, Hefei 230026, China}
\affiliation{Hefei National Laboratory, University of Science and Technology of China, Hefei 230088, China}

\author{Zheng-Wei Zhou}\email{zwzhou@ustc.edu.cn}
%\affiliation{Laboratory of Quantum Information, University of Science and Technology of China, Hefei 230026, China}
\affiliation{Anhui Province Key Laboratory of Quantum Network, University of Science and Technology of China, Hefei 230026, China}
\affiliation{CAS Center For Excellence in Quantum Information and Quantum Physics, University of Science and Technology of China, Hefei 230026, China}
\affiliation{Hefei National Laboratory, University of Science and Technology of China, Hefei 230088, China}
\affiliation{Anhui Center for Fundamental Sciences in Theoretical Physics,
University of Science and Technology of China, Hefei 230026, China}

%\author{Xi-Wang Luo\textsuperscript{*}}%
%\author{Zheng-Wei Zhou\textsuperscript{*}}%
%\affiliation{%
% University of science and technology of China
%}%

\date{\today}% It is always \today, today,

\begin{abstract}
High-fidelity coherent population transfer plays a vital role in the realization of quantum memories. However, population transfer with high performance across a broad frequency range is still challenging due to the finite Rabi coupling strength limited by laser powers. 
%particularly in systems involving atomic ensembles with pronounced decoherence effects. 
Here we propose a novel population-transfer scheme by suturing adiabatic control pulses with each pulse covering certain frequency interval, which are connected in a way that neighboring adiabatic pulses have opposite chirping directions. 
Taking the widely utilized hyperbolic-square-hyperbolic pulse as an example, we demonstrate that rapid and robust population transfer can be achieved. The transfer bandwidth scales linearly with the number of suture pulses while maintaining high fidelity, even at the suture points where adiabaticity breaks down. Crucially, these pulses can be realized by a single laser by means of temporal multiplexing.
%we propose a novel scheme employing reversely chirped parallel adiabatic (SAP) pulses. 
For a given bandwidth, this strategy substantially reduces the operational time which is necessary for on demand read-out and suppressing decoherence effects. Our scheme enables a dramatic increase in multimode storage capacity and paves the way for realizing practical quantum networks.
\end{abstract}

\maketitle

%\tableofcontents

%\section{\label{sec:level1}Introduction}

%\par

\textit{\textcolor{blue}{Introduction}}.---Quantum memories serve as fundamental building blocks for long-distance quantum communication networks and distributed processing nodes within quantum computing architectures~\cite{briegel_quantum_1998,RevModPhys.83.33,de_riedmatten_solid-state_2008,a4,saglamyurek_broadband_2011,a11,a66}. Atomic ensemble-based quantum memories offer distinct advantages by leveraging inherent inhomogeneous broadening to effectively store large number of optical modes with high fidelity and long storage time. Significant progress has been made in various physical platforms such as atomic gases, quantum dots, and rare-earth-ion-doped crystals (REICs)~\cite{duan_long-distance_2001,eisaman_electromagnetically_2005,gao_observation_2012,clausen_quantum_2011,a11,a7,zhong_optically_2015,zhang_experimental_2016,a66}, demonstrating different quantum storage protocols (e.g., controlled reversible inhomogeneous broadening, revival of silenced echo, gradient echo memory, atomic frequency comb (AFC)~\cite{afcpra,a6,a10,a3,businger_non-classical_2022,ortu_storage_2022,a5}, and noiseless photon echo~\cite{ma_elimination_2021,zhou_photonic_2023,liu_millisecond_2025}) which enable multimode quantum storage across temporal, spectral, and spatial domains~\cite{PhysRevLett.128.180501,a5,ma_elimination_2021,liu_millisecond_2025,PhysRevLett.115.070502,sinclair_spectral_2014,humphreys_continuous-variable_2014,a8,a9,parigi_storage_2015,parigi_storage_2015,yang_multiplexed_2018,ortu_storage_2022,teller_solid-state_2025,zhou_photonic_2023}. 
Particularly for REICs systems, remarkable achievements have been made recently, including hour-long coherent optical storage and millisecond-lived photonic quantum storage in integrated devices~\cite{ma_one-hour_2021,liu_millisecond_2025}.

For many atomic ensemble quantum storage protocols, two steps are usually involved to write the optical signal into the memory: the signal is first mapped into collective optical atomic excitations with high optical depth for high efficiency, then, the optical atomic populations are transferred to electron or nuclei spin populations with low decoherence for long storage time~\cite{RevModPhys.83.33,a6}. Note that, one need to convert this process after programmed storage time to produce the signal echo~\cite{PhysRevA.82.042309,a2,PhysRevA.82.042309,tian_reconfiguration_2011,PhysRevLett.129.210501}. 
Therefore, fast and robust population transfer across a broad frequency range plays a critical role in such systems,
%We take the atomic frequency comb spin-wave (AFC-SW) protocol as an example to demonstrate the significance of high fidelity and fast broadband population transfer. 
which are usually implemented by employing a pair of broadband adiabatic control pulses~\cite{PhysRevA.82.042309}. 
%However, previous researches have shown that the population transfer fidelity of these control pulses is typically below 70\%, significantly limiting the overall storage efficiency~\cite{a10,ortu_storage_2022}. 
However, the working bandwidth $W$ of the population transfer, crucially determining the multimode storage capacity of the quantum memory, is limited by the maximum Rabi coupling strength $\Omega$ of the control pulses, which is generally small even for strong laser power due to the weak dipole coupling spin states. Their relation reads $W\sim \Omega^2\tau$ with $\tau$ the pulse operation time. 
Meanwhile, the operation should be fast (i.e., a small $\tau$) to reduce decoherence, which further limit the achievable bandwidth. 
For broadband signals with narrow temporal profiles, this limitation can also degrade the population transfer fidelity, which significantly limits the overall storage efficiency~\cite{a10,ortu_storage_2022}
%This low fidelity primarily stems from insufficient Rabi frequency (\(\Omega\)). According to bandwidth (W) scaling relation (W \(\approx \Omega^{2}t\)): achieving a broad bandwidth W under limited \(\Omega\)  requires a longer operation time \(t\), which scales linearly with W. As a result, the extended operation time exacerbates decoherence effects, further degrading the transfer fidelity~\cite{a9}. 

In this work, we address this issue by proposing a novel population-transfer scheme that significantly increase (decrease) frequency bandwidth (temporal resource consumption). We take the atomic frequency comb-spin-wave (AFC-SW) protocol as an example and show that fast and robust populations transfer with  broadband can be realized by suturing hyperbolic-square-hyperbolic (HSH) adiabatic pulses,
%which integrate multiple adiabatic control pulses,
where each pulse covers a specific frequency interval with neighboring pulses chirped in opposite directions. This design combines the high fidelity characteristic of adiabatic pulses with the low time-resource consumption of parallel methods~\cite{a2,a5}. Over the vast majority of the spectral range covered by the suture adiabatic pulses (SAP), each atom will effectively experience an isolated adiabatic pulse (\textcolor{black}{i.e., $\Omega\tau\gg1$ is maintained}). Although adiabaticity breaks down in the vicinity of the suturing points, we derive analytically the transfer fidelity at the suturing points and demonstrate that near-unity fidelity in these areas could be achieved as well with optimized chirp profiles, which can be viewed as the suturing condition.
The resulting  bandwidth scaling relation ($W \sim n\Omega^{2}\tau$) exhibits linear dependence on the SAP component number $n$. Compared to conventional control schemes, it achieves an $n$-fold increase in $W$ for fixed pulse duration $\tau$, or equivalently, an $n$-fold reduction in $\tau$ for fixed $W$. Our model could be implemented by time-multiplexing of a single laser beam utilizes acoustic-optical-modulator (AOM)~\cite{PhysRevApplied.23.014031} that attains target functionalities without the need for additional laser sources. 

 %\par The paper is organized as follows. Section II reviews the fundamental principles of the AFC-SW protocol. In Section III, we demonstrate our control scheme explicitly and analyze the properties of SAP over adjacent areas. Section IV evaluates the SAP performance through four key aspects: population transfer efficiency; tolerance to pulse duration and amplitude control errors; and achievable bandwidth under practical constraints. Finally, Section V summarizes our work and discusses potential directions for improving quantum memory architectures and high efficiency error-resistant control strategies.

%\section{Introduction to control pulses used in solid state quantum memory}

\begin{figure}
    \centering
    \includegraphics[width=1.0\linewidth]{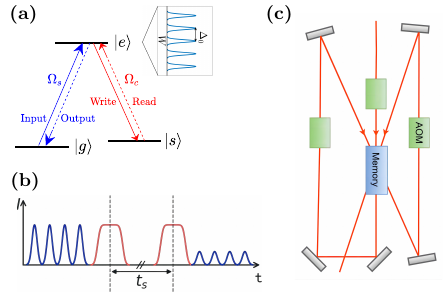}
    \caption{(a) Schematic diagram of AFC-SW protocol: An inhomogeneously broadened transition $\ket{g} \to \ket{e}$ is engineered into frequency combs (periodicity $\Delta_{0}$, bandwidth $W$). Signal pulse $\Omega_s$ excites comb modes, then a pair of control pulses $\Omega_c$ working on $\ket{e} \leftrightarrow \ket{s}$ implement write and read process to emit the signal after certain storage time; (b) AFC-SW protocol illustrated by time-order: Time sequence of an AFC echo memory could be retrieved with certain time interval $t_{s}$ by applying a pair of control pulses;
    (c) Set up of SAP3 : we utilize single laser three times by reflect it twice, using proper AOMs to create three different frequency component of control beams working on the memory and and this scheme could be extended to SAPn.}
    \label{fig1}
\end{figure}

\textit{\textcolor{blue}{AFC-SW protocol}}.---The AFC-SW protocol leverages an inhomogeneously broadened optical transition between atomic ground state $\ket{g}$ and excited state $\ket{e}$, spectrally engineered into a comb structure comprising narrow absorption peaks separated by $\Delta_{0}$ across a broad bandwidth $W$ as depicted in \subfigref{fig1}{a} ~\cite{a6}. When a signal photon resonant with the $\ket{g}-\ket{e}$ transition is absorbed, it excites a delocalized Dicke state
\begin{equation}
    \sum_{j=1}^M e^{-i\Delta_j t} e^{-ik_s z_j} \ket{g_1 \cdots e_j \cdots g_M},
\end{equation}
where $j$ denotes the $j$-th atom in the system, $\Delta_j = m_j \Delta_{0}$ ($m_j \in Z$) reflects the frequency detuning of atom, $z_{j}$ is the position of the atom, $k_{s}$ is the wave number of signal photon. The collective excitations rephase and produce a photon echo at $t_{echo} = 2\pi/\Delta_{0}$, re-emitting the signal photon. This process arises from the rephasing of the initially dephased atomic excitations, a hallmark of coherent collective dynamics in dense atomic ensembles.
The AFC-SW's multimode capacity stems from its spectral-temporal duality: temporal modes of duration $T_{s} \simeq 1/W$ can be stored sequentially within the echo time window $2\pi/\Delta_{0}$, enabling storage of $N_s \simeq 2\pi W/\Delta_{0}$ modes. For long-time storage, as \subfigref{fig1}{b} illustrated, a control pulse with wave number denoted as $k_c$ maps the optical excitation to a spin wave via a metastable state $\ket{s}$, freezing phase evolution for a duration $t_{s}$~\cite{PhysRevA.82.042309}. The spin wave
\begin{equation}
    \sum_{j=1}^M e^{-i\Delta_j \left(t - t_s\right)} e^{-i(k_s - k_c)z_j} \ket{g_1 \cdots s_j \cdots g_M}
\end{equation}
is retrieved using another control pulse, emitting photons backward at a programmable time $2\pi/\Delta_{0}+t_s$ and thus achieve on-demand quantum storage. While conventional $\pi$ pulses demand large Rabi frequency, chirped adiabatic pulses enable efficient spin-wave transfer with much lower Rabi frequency, trading pulse duration for reduced laser power---a critical advance for scalable quantum memory implementations.

%\par In our scheme, as Fig.1(c) shows we utilize three AOM to create three spectrally and spatially separated beams which could be viewed as SAP3 working together.

%\section{Methods}

\textit{\textcolor{blue}{Results}}.---We analyze the three-level atomic system in the AFC-SW protocol, focusing on the coherent coupling between the excited state $\ket{e}$ and spin-wave state $\ket{s}$. The control Hamiltonian takes the general form~\cite{PhysRevA.82.042309}
\begin{equation}
H_{c}(t) = \bar{\Omega}(t)\cos\left[\omega_{0}t - \int_{0}^{t}\Delta(t')d t' + k_c z\right]\sigma_{x},
\end{equation}
where $\bar{\Omega}(t)$ denotes the time-dependent Rabi frequency, $\omega_{0}$ corresponds to the $\ket{e}$ and $\ket{s}$ transition frequency, and $\Delta(t)$ characterizes the chirp profile. We expect such control pulse to realize population transfer from initial state $\ket{\psi(0)} = \ket{e}$ to final state $\ket{s}$. Further, to quantitatively evaluating the pulses' performance, we define the population transfer fidelity as
\begin{equation}
    F = |\bra{\psi(\tau)}\ket{s}|^{2}.
\end{equation}
Previous experimental implementations employed various adiabatic control pulses, including basic $\pi$ pulse, complex hyperbolic-secant pulse, modulated adiabatic pulse, and \textcolor{black}{hyperbolic-square-hyperbolic (HSH) pulse}~\cite{PhysRevA.82.042309,a2,ma_one-hour_2021}. These pulses differ fundamentally in their temporal profiles of $\bar{\Omega}(t)$ and $\Delta(t)$, with HSH pulse demonstrating higher fidelity for robust population transfer across inhomogeneously broadened media. Nevertheless, with limited Rabi frequency or operation time \textcolor{black}{its} effectiveness reduced drastically. 

\textbf{HSH adiabatic pulse.}
The model we used is based on widely used HSH pulses which could be split into three segments: two hyperbolic parts on the edge and one linear part in the middle~\cite{tian_reconfiguration_2011}. Here, in order to satisfy our requirement , we did a little modifications to the edges of HSH pulse which doesn't affect its overall performances. In order to satisfy adiabatic condition, by carefully selecting parameters this pulse could achieve near unity efficiency in broadband population transfer. The parameters of HSH pulse in the interval $[0,\tau]$ are as below: $\Delta(t)$ represents frequency chirp profile, 
\begin{equation}
\Delta(t) =
    \begin{cases} 
  r\times\tanh{\left(\frac{t - t_1}{T}\right)} - \frac{r_1 \times t_2}{2T}  & \text{for } 0\leq t < t_1 \\
 r_1\times(t - t_1 - t_2/2)/T & \text{for } t_1 \leq t \leq t_1 + t_2 \\
   r\times \tanh{\left(\frac{t - t_1 - t_2}{T}\right)}+\frac{r_1\times t_2}{2T} & \text{for } t_1 + t_2<t \leq \tau 
\end{cases},
\end{equation}
while $\Omega(t)$ represents the Rabi frequency of the pulse, 
\begin{eqnarray}
\bar{\Omega}(t) =
\begin{cases} 
\Omega\times\ \sech[\frac{(t-t_1)}{T}] & \text{for } 0\leq t < t_1 \\
\Omega & \text{for } t_1 \leq t \leq t_1 + t_2 \\
\Omega\times \sech[\frac{(t-t_1-t_2)}{T}] & \text{for } t_1 + t_2<t \leq \tau 
\end{cases}.
\end{eqnarray}
Where $T$ depicts the shape of the edge part, $r$ denotes frequency sweeping rate of the edge part while $r_{1}$ denotes frequency sweeping rate of the central linear part, $\Omega$ denotes the maximum Rabi frequency.
However, due to limited $\Omega$, this approach becomes inadequate as it requires long operation time to cover large bandwidth, significantly reducing overall fidelity as a result of decoherence effect~\cite{a5,a9}. To overcome this limitation, we developed a new parallel approach that combines multiple HSH pulses operating simultaneously while attaining high efficiency.

\begin{figure}
    \centering
    \includegraphics[width=1.0\linewidth]{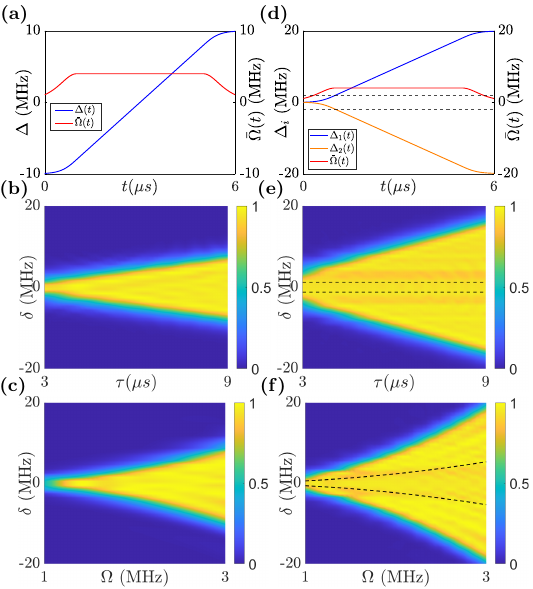}
    \caption{Pulse shape of SAP1 (a) and SAP2 (d), with $\Delta_{1}(t)=\Delta(t)+\frac{f}{2}, \Delta_{2}(t)=-\Delta(t)-\frac{f}{2}$. Two dimensional population transfer fidelity plots that illustrate bandwidth scaling relation with pulse duration $\tau$ and Rabi frequency $\Omega$ of SAP1 (b) (c) and SAP2 (e) (f). We set $\Omega = $ 4 MHz, $t_{1}$ = 1 $\mu s$,  $r = 2$ MHz, $r_{1} = 2$ MHz, $T = 0.5$ $\mu s$ for (a) and (d); \(\Omega = \) 2 MHz, $t_{1}$ = 1 $\mu s$, \textcolor{black}{$r=1.2~$MHz, $r_{1}=0.7~$MHz, $T=0.35~\mu s$} for (b) and (e); $\tau = 6$ $\mu$s, $t_{1}$ = 1 $\mu s$, \textcolor{black}{$r=0.3\times\Omega^{2}, r_{1}=0.15\times\Omega^{2}, T = 0.4~\mu s$} for (c) and (f).}
    \label{fig2}
\end{figure}

\textbf{Suture adiabatic pulses.}
Generally, to cover a broad frequency bandwidth, traditional adiabatic pulses require a duration that increases linearly with the bandwidth when the maximum Rabi frequency is fixed~\cite{ortu_storage_2022,a5}. The suture adiabatic pulses we propose significantly accelerate the population transfer by dividing the full bandwidth into multiple frequency intervals, each covered by a dedicated SAP component. The general form of an $n$-component suture adiabatic pulses' (SAPn) Hamiltonian reads
\begin{equation}
\begin{aligned}
    &H(t) = \frac{(\omega_{1}+\delta)}{2}\sigma_{z}+\bar{\Omega}(t)\sigma_{x}\times\\
    &\sum_{m = 1}^{n}\cos[\omega_{m}t+(-1)^{m}\int_{0}^{t}\Delta(t')d t'+k_c z+\varphi_{m}].
\end{aligned}
\end{equation}
Where $m$ denotes SAPn's $m$-{th} component while $\omega_m=\omega_0-mf+\frac{n+1}{2}f$ with $f$ the frequency range of a single SAP component, $\delta$ represents detuning of the atom and $\varphi_{m}$ is individual phase of each beam. As the individual phase $\varphi_{m}$ is random, our numerical simulations average over it and find its impact to be negligible. In the following analysis, we will set $\varphi_m=0$ for simplicity, the results for nonzero $\varphi_m$ are similar. We could readily interpret the first term of $H(t)$ as an atom's energy level. While the second term is the summation of $n$ adiabatic control pulses where neighboring pulses' center frequency is separated by $f$ and has opposite chirping direction. 
\textcolor{black}{This choice of chirping direction allows one to take advantage of inter-pulse interference at the suture points and the regions near suture points could benefit from this as well (see Appendix A and B for more details).}
A key advantage of this model's physical implementation is that it requires only a single laser beam, which is time-multiplexed to avoid introducing additional laser sources. 
\textcolor{black}{In particular, we reflect the laser beam back through the memory cell after its initial pass, and repeat this process iteratively. Each pass is individually controlled by a high-efficient AOM in order to shape the pulse into its target form. By doing this, we can achieve $n$-fold bandwidth expansion without increasing the total power consumption. The temporal multiplexing introduces a small time delay (typically a few nanoseconds) between different sutured pulses, which is negligible for our purposes.}
\textcolor{black}{Our numerical simulations are based on this model, incorporating individual phase-averaged calculations and accounting for laser power attenuation due to multiple passes through the AOM~\cite{PhysRevApplied.23.014031} by considering a $10\%$ of power decrease which corresponding to a $5\%$ Rabi frequency decrease every time the beam passes through an extra AOM as \subfigref{fig1}{c} shows.} For simplicity, the attenuation coefficient of the Rabi frequency was omitted in following equations, as its effect had a negligible impact on the final population transfer fidelity.
\par Without loss of generality, we move to interaction picture with respect to $H_{0}(t) = \frac{\omega_{1}-\Delta(t) }{2}\sigma_{z}$, making rotating-wave approximation (RWA) because $\omega_{0} \gg \Omega$ and obtain 
\begin{equation}
\begin{aligned}
    &H_{I}(t) = \frac{\delta+\Delta(t)}{2}\sigma_{z} + \frac{\bar{\Omega}(t)}{2}\sigma_{x} + \\
    &\frac{\bar{\Omega}(t)}{2}\sum_{m=2}^{n}[e^{i(m-1)f t-i[1+(-1)^{m}]\cdot\int_{0}^{t}\Delta(t')d t'}\sigma_{+} +h.c. ].
\end{aligned}
\end{equation}
We assume that the atom lies within the frequency interval of the first component of SAPn. Under this assumption, the first two terms of $H_{I}(t)$ can be interpreted as describing a spectrally detuned atom undergoing an adiabatic pulse, while the third term can be treated as noise. Since $f \gg \Omega$ in most cases, the influence of other SAPn components can be neglected under RWA.

\par \textcolor{black}{The optimal performance of SAPn requires specific conditions that is the segmented bandwidth $W/n$ should be much larger than the Rabi frequency $\Omega$, and thus adiabaticity can be assumed to hold in the far-detuned regions away from the suture points, which constitute the majority of the total bandwidth $W$. For $W/n<\Omega$, SAPn's performance could become unreliable because the far-detuned condition breaks down. In such cases, interference between neighboring pulses may degrade the overall performance. Moreover, although a large number of frequency windows is theoretically possible, the scalability of our scheme is practically limited by the aforementioned power attenuation of multiple passes.}

\par The core innovation of our scheme lies in its fulfillment of the suture condition, which is achieved through two essential design principles: first, the waveforms of adjacent adiabatic pulses connect seamlessly at the suture points; second, each pulse exhibits a chirp direction opposite to that of its neighboring pulse. This architecture enables atoms near the suture points to undergo population transfer characterized by damped oscillation,  while those away from suture points experience a full adiabatic transition. As a result, atoms across the entire bandwidth can achieve population transfer fidelity close to unity. A key advantage of this sutured design is its robustness against various control errors, significantly enhancing operational reliability and fidelity across the entire bandwidth.

\begin{figure}
    \centering
    \includegraphics[width=1.0\linewidth]{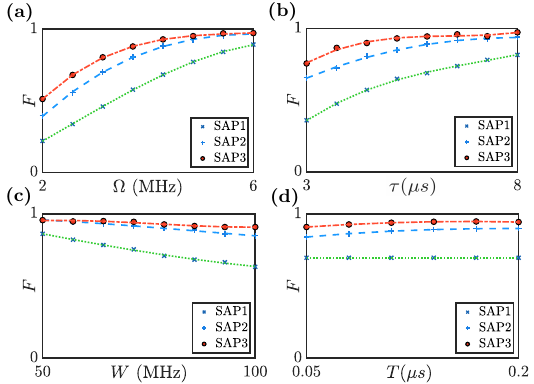}
    \caption{Comparison of population transfer fidelity $F$ for three control pulses across four key parameters with fixed $t_{1}$ = 0.5 $\mu s$ while other parameters ($T,r,r_{1}$) optimized for (a)(b)(c) and ($r,r_{1}$) optimized for (d). (a) $F$ as a function of maximum Rabi frequency $\Omega$ (fixed $\tau$ = 5 $\mu s$, $W$ = 100 MHz); (b) $F$ as a function of pulse duration $\tau$ (fixed \(\Omega =\) 4 MHz, $W$ = 100 MHz); (c) $F$ as a function of bandwidth $W$ (fixed $\Omega =$ 4 MHz, $\tau$ =  5 $\mu s$); (d) $F$ as a function of pulse shape parameter $T$ ( fixed $\Omega =$ 4 MHz, $\tau$ = 6 $\mu s$, $W$ = 100 MHz).}
    \label{fig3}
\end{figure}

\textbf{Two component suture adiabatic pulses.}
To analyze the property of suture adiabatic pulses, its straightforward to start with two component suture adiabatic pulses and based on this we could easily extend to multi-component situations. The Hamiltonian of SAP2 reads
\begin{equation}
\begin{split}
    H^{\prime}(t) = &\frac{(\omega_{0}+\delta)}{2}\sigma_{z}+\bar{\Omega}(t)\cos[\omega_{0}t-\frac{f t}{2}+\int_{0}^{t}\Delta(t')d t']\sigma_{x}\\
    &+\bar{\Omega}(t)\cos[\omega_{0}t+\frac{f t}{2}-\int_{0}^{t}\Delta(t')d t']\sigma_{x}.
\end{split}
\end{equation}
It is reasonable for us to assume that the two component suture adiabatic pulses could work separately as individual adiabatic pulse in vast majority frequency range except for their adjacent areas. 
%\textcolor{black}{Note that the adiabatic condition requires $\Omega^2\tau$ much larger than $f$, so the constrain $\Omega\tau\gg1$ is maintained.}
We deliver a specified illustration of SAP1 and SAP2's pulse shape within the vicinity of the suture point being outlined particularly by two black dashed lines in \subfigref{fig2}{a} and \subfigref{fig2}{d}. These areas could be interpreted as the hyperbolic edge parts of adjacent SAP components which the spectral distance between two lines could be denoted as $2\times[\Delta(t_{1})- \Delta(0) ] $. Further, we deliver four two dimensional population transfer fidelity plots with parameters ($T, r, r_{1}$) optimized to characterize the bandwidth scaling properties of SAP1 and SAP2. The results reveal a twofold enhancement in bandwidth achieved by SAP2 over SAP1, both under identical Rabi frequency (\subfigref{fig2}{b}, \subfigref{fig2}{e}) and identical pulse durations  (\subfigref{fig2}{c}, \subfigref{fig2}{f}). \textcolor{black}{The inhomogeneity visible in \subfigref{fig2}{b} and \subfigref{fig2}{c} arises because SAP1 is an adiabatic pulse of finite duration and is therefore not strictly adiabatic over the full bandwidth. This explanation also applies to the inhomogeneity in \subfigref{fig2}{e} and \subfigref{fig2}{f}. For SAP2, the inhomogeneity is slightly enhanced around the dashed lines, which is due to inter-pulse interference, though its contribution to the total fidelity loss remains minor.} We move to interaction picture with respect to $H_{0}^{\prime} = \frac{\omega_{0}}{2}\sigma_{z}$
\begin{equation}
\begin{split}
    H^{\prime}_{I}(t) = &\frac{\delta}{2}\sigma_{z}+\frac{\bar{\Omega}(t)}{2}[e^{-i[\frac{f t}{2} - \int_{0}^{t}\Delta(t')d t']}\sigma_{+} + h.c.] \\
    &+ \frac{\bar{\Omega}(t)}{2}[e^{i[\frac{f t}{2} - \int_{0}^{t}\Delta(t')d t']}\sigma_{+} + h.c.].
\end{split}
\end{equation}

\par its obvious that this Hamiltonian could be written in a more compact form
\begin{equation}
\begin{split}
    H^{\prime}_{I}(t) = &\frac{\delta}{2}\sigma_{z}+\bar{\Omega}(t)\cos[\frac{f t}{2} - \int_{0}^{t} \Delta(t^{\prime})dt^{\prime}]\sigma_{x} .
\end{split}
\end{equation}
This Hamiltonian can be interpreted as a frequency-detuned atom interacting with a complex oscillating $\pi$ pulse. We focus mainly on situations where $\delta$ is small. In this case, we could assume that the atom experiences a time-dependent $\pi$ pulse with damped oscillation. With simple optimization of pulse's parameters we could achieve near-unity efficiency around the adjacent areas between two pulses. We deliver detailed discussions about this property in our \textcolor{black}{Appendix A}.

\begin{figure}
    \centering
    \includegraphics[width=1.0\linewidth]{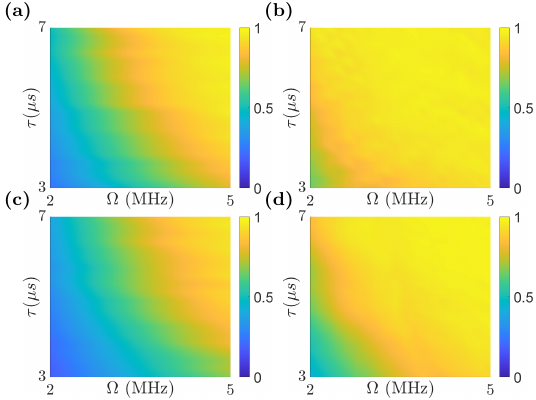}
    \caption{Contour plots of population transfer fidelity $F$ as a function of maximum Rabi frequency \(\Omega\) and the pulse duration $\tau$ over bandwidth of 50 MHz in (a) (b) and 80 MHz in (c) (d), with
    control protocols SAP1 in (a) (c) and SAP3 in (b) (d). The parameters ($T, r, r_{1}$) are optimized with fixed $t_{1}$ = 0.5 $\mu$s.}
    \label{fig4}
\end{figure}

\textbf{Performance evaluation.} There are several metrics to evaluate an adiabatic pulse's performance. Among which, population transfer fidelity stands out as major factor that determines whether an adiabatic pulse is applicable. Several experiments carried out before had their overall storage efficiency severely diminished due to rather low efficiency of control pulses~\cite{ortu_storage_2022,businger_non-classical_2022}. This property could be even more destructive when the storage scheme has to deal with more control pulses. Achievable maximum  Rabi frequency $\Omega$ and the duration of control pulse $\tau$ are two main parameters of its population transfer fidelity. Here, we investigate population transfer fidelity in a two-level system across varying $\Omega$ and $\tau$ in \subfigref{fig3}{a} and \subfigref{fig3}{b}.  The multi-component SAP pulse significantly
outperforms the SAP1 over a wide range of
both $\Omega$ and $\tau$. We also examine their performance under different bandwidth conditions and pulse shape respectively in \subfigref{fig3}{c} and \subfigref{fig3}{d}. Multi-component SAP pulse excels especially in large bandwidth region comparing with SAP1. Meanwhile pulse shape distortion hardly contribute to population transfer efficiency diminishment.

\begin{figure}
    \centering
    \includegraphics[width=1.0\linewidth]{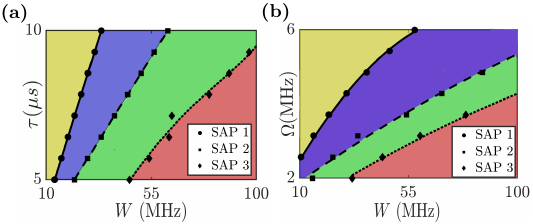}
    \caption{(a) Phase diagram depicting population transfer fidelity $F$ near 0.95 across three distinct pulses as a function of bandwidth $W$ and pulse duration $\tau$ with fixed maximum Rabi frequency $\Omega$= 3 MHz and $t_{1}$ = 0.5$\mu s$; (b) Phase diagram depicting population transfer fidelity $F$ near 0.95 across three distinct pulses as a function of bandwidth $W$ and maximum Rabi frequency $\Omega$ with fixed pulse duration $\tau$= 5 $\mu$s and $t_{1}$ = 0.5$\mu s$. Other parameters ($T, r, r_{1}$) have been optimized.}
    \label{fig5}

\end{figure}

\textbf{Imperfection effects.}
In realistic experimental implementations, various sources of control imperfections can introduce non-negligible attenuation in population transfer fidelity. Among these, pulse duration errors and Rabi frequency deviations are particularly significant, as they directly perturb the coherent dynamics of the system. To quantitatively assess the robustness of different pulse schemes against these errors, as illustrated in \subfigref{fig4}{a}–\subfigref{fig4}{d}, we perform numerical simulations across a broad parameter space with $\Omega$ and $\tau$ ranging from 2 MHz to 5 MHz, and 3 $\mu$s to 7 $\mu$s, respectively. And we demonstrate that SAP1 showing pronounced sensitivity to deviations in both pulse duration and Rabi frequency while SAP3 is much more resilient against these errors. This enhanced robustness can be attributed to multi component SAP's rather slow frequency sweeping rate comparing with SAP1. \textcolor{black}{Effects of other  practical imperfections are discussed in Appendix C.}

\textbf{Achievable band-width.}
The constrained bandwidth presents a fundamental limitation on multimode capacity. As illustrated before, multimode storage capacity $N_s$ scales linearly with bandwidth $W$. To systematically investigate this bandwidth-dependent behavior, we conducted a comprehensive phase diagram analysis of population transfer fidelity across three distinct pulse types under varying bandwidth and pulse duration (maximum Rabi frequency) conditions in \subfigref{fig5}{a} (\subfigref{fig5}{b}). We set the population transfer fidelity threshold at 0.95 to investigate different pulses' best working domains. Our findings demonstrate that multi-component SAP exhibit superior performance characteristics in the regime of increased bandwidth and reduced pulse duration. The enhanced performance of multi-component SAP under these conditions shows great potential for highly efficient multimode quantum storage.

\textit{\textcolor{blue}{Conclusion}}.---The model we propose offers a promising solution for large bandwidth quantum storage~\cite{businger_non-classical_2022,ortu_simultaneous_2018}. By leveraging novel control pulse designs, our approach effectively mitigates the inherent trade-offs in conventional methods. It not only preserves the advantages of broadband and uniform population transfer but also enhances storage bandwidth, reduces pulse duration, and improves overall storage efficiency. This makes it a significant step forward in addressing the pressing challenges in solid-state quantum memory systems and paves the way for broader applications in quantum information science.
The versatility of our approach makes it applicable to a large variety of quantum technologies. For instance, it holds potential for enhancing quantum sensing by enabling high efficiency control and measurement over a wide range of frequencies. Similarly, in the context of quantum control based on ensembles of atoms, our method could preserve systems' coherence by shorten control time while maintaining control fidelity, thereby enhancing the performance of protocols reliant on collective quantum states. Our method can also be applied in the field of dynamical decoupling, where precise and efficient control over quantum states is critical for mitigating the effects of decoherence.
Furthermore, the techniques we discuss could be adapted to other types of quantum memory systems, broadening the scope of their applicability in diverse architectures and platforms.

\begin{acknowledgments}
We thank Dr. Zong-Quan Zhou for fruitful discussions. This work was funded by the National Natural Science Foundation of China (Grants No. 12474366 and No. 12574544) and Innovation Program for Quantum Science and Technology (Grant No. 2021ZD0301200).
X.-W. Luo also acknowledges the support from USTC start-up funding.
\end{acknowledgments}

\appendix

\section{Appendix A: Efficiency evaluation of suture points}

\begin{figure}
    \centering
    \includegraphics[width=0.8\linewidth]{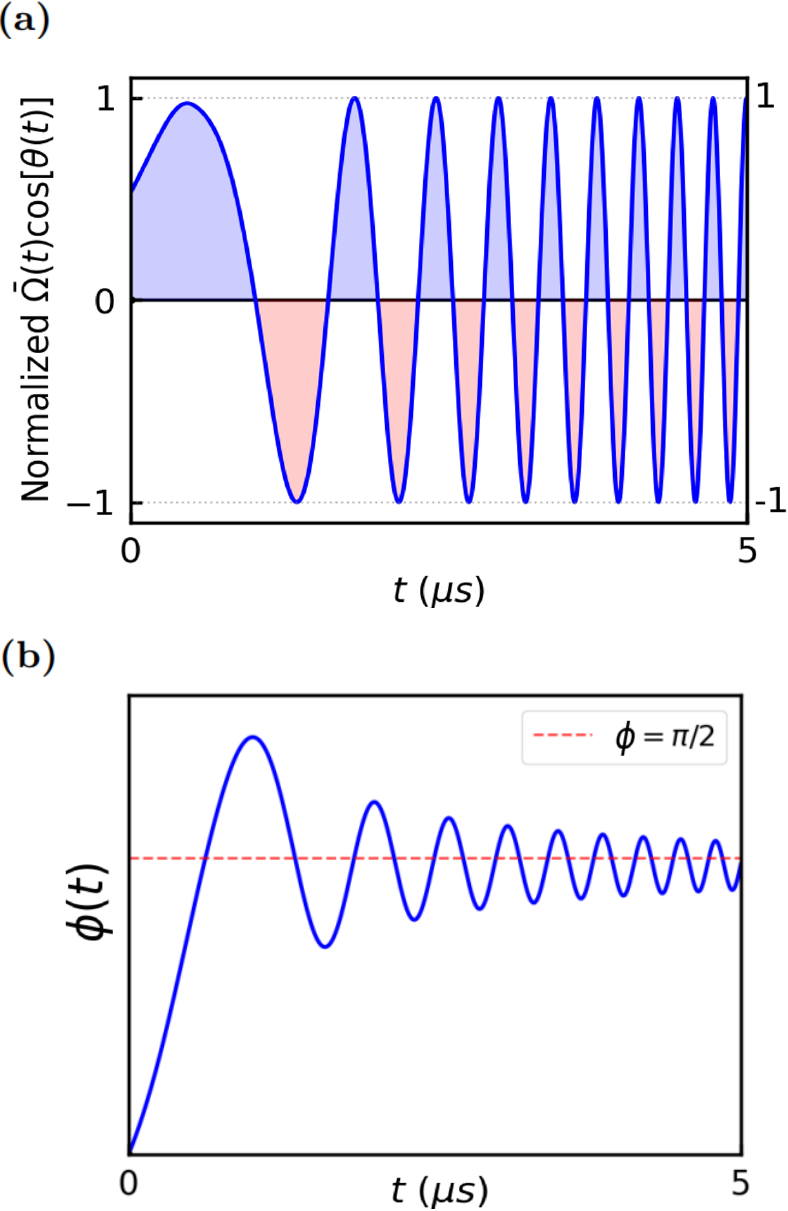}
    \caption{(a) The plot depicts the oscillatory behavior of $\bar{\Omega}(t)\cos[\theta(t)]$, where blue regions indicate monotonically increasing segments in (b); while red regions denote monotonically decreasing segments in (b). The plot demonstrates progressively decreasing oscillation period and amplitude of $\phi(t)$ over time. Here, we set $t_{1}$ = 0.5 $\mu s$, $T$ = 0.4 $\mu s$, $\Omega$ = 3 MHz, $r$ = 1.5 MHz, $r_{1}$ = 2 MHz. }
    \label{fig.A.1}
\end{figure}

The Hamiltonian of two component suture adiabatic pulses reads
\begin{equation}
\begin{split}
    H^{\prime}_{I}(t) = &\frac{\delta}{2}\sigma_{z}+\bar{\Omega}(t)\cos[\frac{f t}{2} - \int_{0}^{t} \Delta(t^{\prime})dt^{\prime}]\sigma_{x} 
\end{split}.
\end{equation}
We consider the exact adjacent point where $\delta = 0$,
\begin{equation}
\begin{split}
    H^{\prime}_{I}(t) = &\bar{\Omega}(t)\cos[\frac{f t}{2} - \int_{0}^{t} \Delta(t^{\prime})dt^{\prime}]\sigma_{x} 
\end{split}.
\end{equation}
Defining the phase accumulation function $\theta(t) = \frac{f t}{2} - \int_{0}^{t} \Delta(t^{\prime})dt^{\prime} $ yields
\begin{equation}
\begin{split}
    H^{\prime}_{I}(t) = &\bar{\Omega}(t)\cos[\theta(t)]\sigma_{x} .
\end{split}
\end{equation}

\par Thus, the problem has been transformed into analyzing a complicated time-dependent $\pi$ pulse's property. Initial state is defined as $\ket{\psi(0)} = \ket{e}$, we can decompose this state into $\frac{\ket{+}+\ket{-}}{\sqrt{2}} $, where $\ket{+} = \frac{\ket{e}+\ket{s}}{\sqrt{2}}$ and $\ket{-} = \frac{\ket{e}-\ket{s}}{\sqrt{2}}$. Under the evolution governed by the Hamiltonian described above, the final state is:
\begin{equation}
\begin{split}
    \ket{\psi(\tau)} = &\frac{e^{-i\int_{0}^{\tau}\bar{\Omega}(t)\cos[\theta(t)]dt}\ket{+}+e^{i\int_{0}^{\tau}\bar{\Omega}(t)\cos[\theta(t)]dt}\ket{-}}{\sqrt{2}} .
\end{split}
\end{equation}
For simplicity, we denote $\int_{0}^{t}\bar{\Omega}(t^{\prime})\cos[\theta(t^{\prime})]d t^{\prime}$ as $\phi(t)$. The fidelity $F$ with respect to the target state $\ket{s}$ is defined as:
\begin{equation}
\begin{split}
    F = |\bra{\psi(\tau)}\ket{s}|^{2} = \sin^{2}[\phi(\tau)].
\end{split}
\end{equation}
Apparently, setting $\phi(\tau) = \frac{\pi}{2}$ at the end of evolution is equivalent to attaining unity population transfer fidelity. Since $\bar{\Omega}(t)\cos[\theta(t)]$ has oscillation property, as \subfigref{fig.A.1}{b} illustrates, $\phi(t)$ also exhibits oscillatory behavior. Clearly, as \subfigref{fig.A.1}{a} depicts, when $
\theta(t) = \frac{\pi}{2} + n\pi \quad (n \in Z),
$ $\cos[\theta(t)]$ changes sign and the monotonicity of $\phi(t)$ changes. Furthermore, we make an approximation that the areas $\bar{\Omega}(t)\cos[\theta(t)]$ encircles with x-axis when $\theta(t)>\frac{\pi}{2}$ could be treated as triangles, so the variation magnitude of $\phi(t)$ scales proportionally to the temporal length of its monotonic interval. Given the first maximum $\phi(t_0)$ of $\phi(t)$ with $t_{0}>t_{1}$, for simplicity, we denote $a = \frac{r_{1}}{2T}$, $b = \frac{[\frac{f}{2}+\Delta(t_{0})] }{2a}$. The function takes the form:
\begin{equation}
\begin{split}
    &\phi(t) =  \phi(t_{0}) + \frac{\Omega}{2} (\sqrt{b^{2}+\frac{\pi}{a}}-b)\\&+\frac{\Omega}{2}\sum_{k=2}^{n}(-1)^{k}[\sqrt{b^{2}+\frac{k\pi}{a}}-\sqrt{b^{2}+\frac{(k-1)\pi}{a}}].
\end{split}
\end{equation}

By the Leibniz convergence criterion for alternating series, $\phi(t)$ approaches a steady-state value, implying fidelity stabilization under extended pulse interaction. By parameter optimization, we could always achieve near unity fidelity.

\begin{figure}
    \centering
    \includegraphics[width=0.99\linewidth]{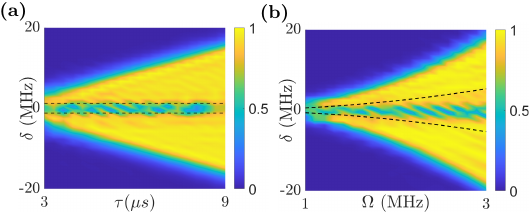}
    \caption{\textcolor{black}{Two dimensional population transfer fidelity plots with SAP2 but two pulses share the same chirp direction. (a) Contour plot of duration $\tau$; (b) Contour plot of Rabi frequency $\Omega$. We set \(\Omega = \) 2 MHz, $t_{1}$ = 1 $\mu s$, $r=1.2~$MHz, $r_{1}=0.7~$MHz, $T=0.35~\mu s$ for (a); $\tau = 6$ $\mu$s, $t_{1}$ = 1 $\mu s$, $r=0.3\times\Omega^{2}, r_{1}=0.15\times\Omega^{2}, T = 0.4~\mu s$ for (b). }}
    \label{fig7}
\end{figure}

\textcolor{black}{\section{Appendix B: Effects of chirping direction}}

\textcolor{black}{The opposite chirp direction between neighboring adiabatic pulses is designed to utilize inter-pulse interference constructively at the suture points. This approach ensures that the regions near these points benefit from interference, while adiabaticity is maintained across most of the bandwidth where interference is negligible, thereby securing high fidelity throughout the total bandwidth.   }
\par\textcolor{black}{We have conducted a detailed numerical simulation to evaluate the system performance under the condition where all pulses share the same chirp direction in \subfigref{fig7}{a} and \subfigref{fig7}{b} with same parameters comparing with \subfigref{fig2}{e} and \subfigref{fig2}{f} respectively. The result reveals a significant degradation in fidelity. This performance deterioration, is particularly pronounced in the spectral adjacent areas between neighboring pulses, suggesting that destructive interference effect between neighboring pulses arise when they are identically-chirped.}

\begin{figure}
    \centering
    \includegraphics[width=0.99\linewidth]{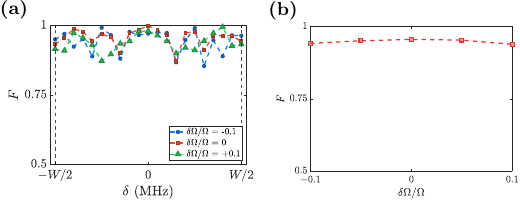}
    \caption{\textcolor{black}{(a) Fidelity in the presence of Rabi frequency error $\delta\Omega$ for SAP2. We select three Rabi frequency values $\delta\Omega/\Omega=[-0.1,0,0.1]$.  (b) Bandwidth averaged fidelity in the presence of Rabi frequency error for SAP2. The Rabi frequency error varied from $\delta\Omega/\Omega=-0.1$ to 0.1. The parameters are set as follows: $W = 20$~MHz, $\Omega = 3$~MHz, $t_{1}=0.5\mu s, t_{2}=5\mu s$ with other parameters ($T,r,r_{1}$) optimized at suture point for $\delta\Omega=0$. We consider $\varphi_m=0$ for all $m$. }}
    \label{fig8}
\end{figure}

\textcolor{black}{\section{Appendix C: Effects of practical imperfections}}
\textcolor{black}{In realistic experiments, multiple passes of the laser through the memory cell (i.e., temporal multiplexing) may reduce the Rabi frequency $\Omega$, which present a practical limitation to the scalability of our scheme. Separately, experimental fidelity is compromised by several real-world factors. Within the framework of our theoretical model, laser frequency noise and inter-pulse interference emerge as notable contributors to the degradation of transfer fidelity. Furthermore, the fluctuation of Rabi frequency also leads to efficiency reduction under typical operating conditions. }
\par\textcolor{black}{ To systematically quantify the impact of these effects, we have conducted a series of targeted numerical simulations designed to evaluate their influence. 
Specifically, we deliver a numerical simulation to investigate Rabi frequency error in \subfigref{fig8}{a} and \subfigref{fig8}{b}. We find that the suture condition and fidelity are quite robust against the Rabi frequency error. Due to the fast oscillation of the coupling field $\bar{\Omega}\cos[\theta(t)]$, the phase $\phi(\tau)$ will always converge to about $\pi/2$ at the end of the adiabatic process, though it depends linearly on the Rabi frequency. Therefore, the phase $\phi(\tau)$
    only fluctuates slightly around $\pi/2$ (i.e., $\phi(\tau)=\pi/2+\delta\phi$) for different parameters, while the suture point fidelity  $F=\sin^2(\phi)\sim 1-\delta\phi^2$ is dominated by second-order corrections. Moreover, the exact suture condition depends slightly on the relative phase $\varphi_m-\varphi_{m+1}$ between two adjacent pulses, so the suture condition is approximately satisfied for different relative phases. When averaged over the relative phase, the effects of Rabi frequency error is further suppressed.
    }

\begin{figure}
    \centering
    \includegraphics[width=0.99\linewidth]{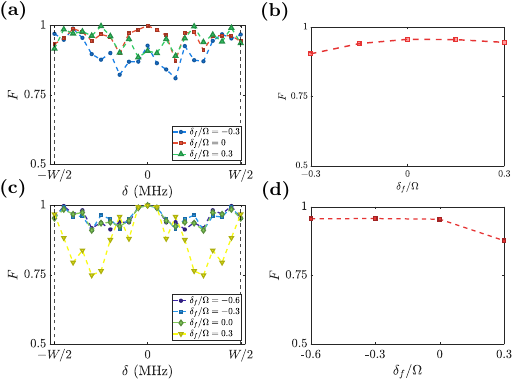}
    \caption{\textcolor{black}{(a) and (c) Fidelity $F$ in the presence of central frequency shift $\delta_f$ for SAP2. $\delta_f<0$ ($\delta_f>0$) corresponds to overlapped (separated) frequency windows of the two pulses. (b) and (d) Bandwidth-averaged fidelity for as a function of central frequency shift $\delta_f$. The parameters are set as follows: $W = 20$~MHz, $\Omega = 3$~MHz, $t_{1}=0.5\mu s, t_{2}=5\mu s$. In (a) and (b), the parameters ($T,r,r_{1}$) maximize the suture-point fidelity for perfect connection $\delta_f=0$. While in (c) and (d), the parameters ($T,r,r_{1}$) are optimized separately for different $\delta_f$. We consider $\varphi_m=0$ for all $m$. }}
    \label{fig:fig9}
\end{figure}

\textcolor{black}{Furthermore, we numerically simulated the effects of central frequency-shift error of each pulse in \subfigref{fig:fig9}{a-d}.
We consider an frequency shift $f\rightarrow f+\delta_f$, such that $\delta_f<0$ ($\delta_f>0$) corresponds to overlapped (separated) frequency windows of the two adjacent pulses.
We consider optimized parameters that maximize the suture-point fidelity for perfect connection, and find that the frequency shift $\delta_f$ breaks the suture condition and lower the fidelity around the suture point. However, the fidelity away from the suture point is hardly affected and may even be slightly increased, the bandwidth-averaged fidelity only slightly decreases. This indicates that our scheme is robust against such frequency fluctuations.
    On the other hand, if we introduce frequency overlap on purpose and re-optimize the parameters for the suture point, then the overlapped case and the connected case lead to almost the same averaged fidelity. We find that the separated case have a lower averaged fidelity, since parameters maximizing the suture-point fidelity do not naturally optimize the bandwidth-averaged fidelity.
    It is worthy to note that the suture condition depends on the relative phase $\varphi_m-\varphi_{m+1}$ of adjacent pulses. As a result, if the parameters are optimized for the fidelity averaged over the bandwidth and relative phase, the overlapped and connected cases achieve nearly identical performance. However, the frequency overlapping may slightly decrease the overall frequency bandwidth.
    }

\nocite{*}

%\bibliography{AFC.bib}% Produces the bibliography via BibTeX.
\providecommand{\noopsort}[1]{}\providecommand{\singleletter}[1]{#1}%

\end{document}